\documentclass[12pt,a4paper]{article}

\newcommand{\pa}{\partial}
\newcommand{\oa}{\tilde{\omega}_0}
\newcommand{\ob}{\tilde{\omega}_1}
\newcommand{\ua}{\tilde{u}_0}
\newcommand{\ub}{\tilde{u}_1}
\newcommand{\si}{\tilde{\sigma}}
\newcommand{\ta}{\tilde{\tau}}
\newcommand{\lu}{\tilde{U}}
\newcommand{\sn}{\mathrm{sn}}
\newcommand{\cn}{\mathrm{cn}}
\newcommand{\dn}{\mathrm{dn}}

\begin{document}

\begin{flushright}
{ }
\end{flushright}
\vspace{1.8cm}

\begin{center}
 \textbf{\Large Conformal SO(2,4) Transformations \\
for the Helical AdS String Solution }
\end{center}
\vspace{1.6cm}
\begin{center}
 Shijong Ryang
\end{center}

\begin{center}
\textit{Department of Physics \\ Kyoto Prefectural University of Medicine
\\ Taishogun, Kyoto 603-8334 Japan}
\par
\texttt{ryang@koto.kpu-m.ac.jp}
\end{center}
\vspace{2.8cm}
\begin{abstract}
By applying the conformal SO(2,4) transformations to the folded rotating
string configuration with two spins given by a certain limit from the 
helical string solution in $AdS_3 \times S^1$, we construct new string
solutions whose energy-spin relations are characterized by the boost
parameter. When two SO(2,4) transformations are performed with two
boost parameters suitably chosen, the straight folded 
rotating string solution with one spin in $AdS_3$ is transformed in 
the long string limit into the long spiky string solution whose 
expression is given from the helical string solution in $AdS_3$ 
by making a limit that the modulus parameter becomes unity.
\end{abstract}
\vspace{3cm}
\begin{flushleft}
March, 2008 
\end{flushleft}

\newpage
\section{Introduction}

The AdS/CFT correspondence \cite{MGW} has more and more revealed the deep
relations between the $\mathcal{N}=4$ super Yang-Mills (SYM) theory
and the string theory in $AdS_5 \times S^5$, where various types of
classical string solutions play an important role. The energy spectrum of
certain string states matches with the spectrum of dimensions of field
theory operators in the SYM theory \cite{BMN,GKP}. There has been
a mounting evidence that the spectrum of AdS/CFT is described by
studying the multi-spin folded or circular rotating string solutions
in $AdS_5 \times S^5$  in a particular large spin limit
\cite{FT,JM,AT} and by analyzing the Bethe equation
for the diagonalization of the integrable spin chain in the planar SYM
theory \cite{MZ,BMS,BFS}. 

The other class of string solutions in $AdS_5$ have been constructed to
have a number of spikes on the string and be associated with higher twist
operators in SYM \cite{MK}. The spiky string solution has been 
generalized to a string configuration in sphere \cite{SR}. By the 
T-duality transformation $\tau \leftrightarrow \sigma$ of the spiky string
solutions in $AdS_5$ and $R \times S^5$ new dual  spiky string solutions 
have been produced \cite{IK,MS}.

In another large spin limit such that both the spin chain and the string 
effectively become very large there has been a construction of a rotating
open string solution with one spin in $R \times S^2$, namely, the giant 
magnon \cite{HM} which is a particular case of the spiky string
in $R \times S^2$ \cite{SR}. The one-spin giant magnon has been identified
with an elementary magnon excitation in the long spin chain where the
energy-spin relation agrees with the strong 't Hooft coupling limit
of that for the spin chain magnon derived from the $SU(2|2)\times SU(2|2)$
supersymmetry with a novel central extension \cite{NB}.
The scattering of two giant magnons has been analyzed such that the giant
magnon is identified with the sine-Gordon soliton. 
By exploiting the equivalence between the string theory in $R \times S^3$
and the complex sine-Gordon theory via Pohlmeyer's reduction \cite{KP}
the dispersion relation for the two-charge dyonic giant magnon has been
constructed \cite{ND}. There have been various investigations such as
the multi-spin giant magnons \cite{AFZ,MTT,SRA,SV,KRA}, the S-matrix
of giant magnons \cite{CDO}, the finite size corrections to the giant 
magnon dispersion relations \cite{AFZ,AFG} and  the semiclassical 
quantization of giant magnons \cite{JAM}. 

Through the equivalence between the O(4) string sigma model
and the complex sine-Gordon model, a more general string  solution  
in $R \times S^3$, namely, the helical string has been constructed 
\cite{OS} in terms of the elliptic theta functions and two parameters,
 the soliton velocity $v$ and the elliptic modulus parameter $k$, to
interpolate between the two-spin folded/circular string in the
$v \rightarrow 0$ limit and the dyonic 
giant magnon in the $k \rightarrow 1$
limit \cite{ND}. There are two kinds of helical strings, the 
type (i) helical string characterized by the number of spikes and
the type (ii) one by the number of crossing the equator of sphere.
The helical string solution has been reconstructed \cite{BV} as a
finite-gap solution \cite{KMM} in the algebro-geometric approach to the
string equations of motion \cite{DV}. 

By perfoming the T-duality transformation $\tau \leftrightarrow \sigma$
to this helical string in $R \times S^3$, a new general string solution
has been presented \cite{HOS} to interpolate between the pulsating
string and the dual spiky string with single spike \cite{IK}.
From the type (i) helical string in $R \times S^3$, the type (iii) helical
string in $AdS_3 \times S^1$ with two spins has been produced by the
analytic continuation of the elliptic modulus squared, while 
the type (iv) helical string has been generated by shifting the boosted
worldsheet space coordinate. The type (iii) helical string becomes
the folded rotating string as well as
 the spiky string \cite{MK} in certain
parameter limits, and the type (iv) helical string includes the SL(2)
giant magnon \cite{MTT,SRA} in the $k \rightarrow 1$ limit.
The finite-size correction to the dyonic giant magnon has been
computed by analyzing the asymptotic behavior of the type (i) helical 
string with two spins in $R \times S^3$ \cite{HS}.

On the other hand in order to study the planar 4-gluon amplitude at 
strong coupling in the SYM theory the open string solution in $AdS_5$ with
Euclidean worldsheet has been constructed \cite{AM} 
for computing the Wilson loop 
with 4 cusps in the T-dual coordinates whose boundary conditions are
determined by the massless gluon momenta. The 4-cusp Wilson loop surface
is related by a certain conformal SO(2,4) transformation to the 
1-cusp Wilson loop surface found in \cite {MKR}.
There have been various investigations of the Euclidean open string 
solutions in $AdS_5$ associated with the planar gluon amplitudes
\cite{AFK,ADI,MMT}.

In ref. \cite{KRT} starting from the long string limit of the folded 
rotating closed string solution with Minkowski 
worldsheet in $AdS_5$ \cite{GKP}, 
the analytic continuation has been taken to yield an open string solution
with Euclidean worldsheet, which is further transformed by a discrete 
SO(2,4) rotation into the 4-cusp Wilson loop solution \cite{AM}.
The 1-loop correction \cite{KRT} and the 2-loop correction without
 spin $J$ in $S^5$ \cite{RT} and with spin $J$ \cite{RRT} to the classical
string solution associated with the 1-cusp Wilson loop surface
have been computed, where the cusp anomaly function is derived
in the strong-coupling expansion. By taking the infinite spin limit
of the spiky rotating string in $AdS_3$, the spiky solution 
has been described in a simple analytic expression \cite{KAT},
where a generic arc of the spiky string connecting two spikes reaching
the AdS boundary is related by the SO(2,4) boosts to the infinite spin
limit \cite{FT,FTT} of the straight folded rotating string \cite{GKP}.
The SO(2) rotated 1-cusp Wilson loop solution in $AdS_5$ 
\cite{KRT,MKR} has been generalized to the case in $AdS_5 \times S^5$
\cite{JK}. There has been another study to generate classical string
solutions in $AdS_5$ by using the Pohlmeyer reduction, where the
4-cusp Wilson loop solution associated with the long string limit of
the straight folded rotating closed string is related to 
the sinh-Gordon vacuum \cite{JJK}.

We will consider the SO(2,4) transformations of the type (iii) helical
string solution in $AdS_3 \times S^1$. The SO(2,4) transformation with
an arbitrary boost parameter will be applied to the straight folded
rotating string in $AdS_3 \times S^1$ which is given by a
particular limit of the type (iii) helical string solution.
The generated string configuration will be shown to satisfy the string
equations of motion and its energy-spin relation will be derived.
We will perform particular SO(2,4) boosts to the long string limit
of the spiky string solution in $AdS_3$ whose expression is
produced from the type (iii) helical string. The SO(2,4) boosted string 
configuration will be shown to have relation with the 4-cusp
Wilson loop solution.
 
\section{SO(2,4) transformations of the folded rotating string}

We consider a closed string in $AdS_3 \times S^1$ whose embedding 
coordinates $\eta_0, \eta_1$ and $\xi_1$ are expressed as
\begin{eqnarray}
\eta_0 &=& Y_0 + i Y_{-1} = \cosh \rho e^{it}, \hspace{1cm}
\eta_1 = Y_1 + i Y_2 = \sinh \rho e^{i\phi_1},  \nonumber \\
\xi_1 &=& X_1 + i X_2 = e^{i\varphi_1}
\end{eqnarray}
and obey
\begin{eqnarray}
\vec{\eta}^*\cdot \vec{\eta} &\equiv& - |\eta_0|^2 + |\eta_1|^2 
= - Y_{-1}^2 - Y_0^2 + Y_1^2 + Y_2^2 = -1, \nonumber \\
|\xi_1|^2 &=& X_1^2 + X_2^2 = 1.
\end{eqnarray}
The Polyakov action becomes
\begin{equation}
S = - \frac{\sqrt{\lambda}}{4\pi} \int d\sigma d\tau \left[ \gamma^{ab}
( \pa_a\vec{\eta}^*\cdot \pa_b\vec{\eta} + \pa_a\xi_1^* \pa_b\xi_1 )
+ \tilde{\Lambda}(\vec{\eta}^*\cdot \vec{\eta} + 1 )
+ \Lambda(\xi_1^* \xi_1 - 1 ) \right],
\label{pac}\end{equation}
from which the equations of motion in the conformal gauge are 
provided by
\begin{equation}
\pa_a\pa^a \vec{\eta} - ( \pa_a\vec{\eta}^*\cdot \pa^a\vec{\eta})
\vec{\eta} = 0, \hspace{1cm}
\pa_a\pa^a \xi_1 + ( \pa_a\xi_1^* \pa^a\xi_1)\xi_1 = 0
\label{st}\end{equation}
and the Virasoro constraints are given by
\begin{eqnarray}
0 &=& T_{\sigma \sigma} = T_{\tau \tau} = \frac{\delta^{ab}}{2}
(\pa_a\vec{\eta}^*\cdot \pa_b\vec{\eta} + \pa_a\xi_1^* \pa_b\xi_1),
\nonumber \\
0 &=& T_{\tau \sigma} = \mathrm{Re}(\pa_{\tau}\vec{\eta}^*\cdot 
\pa_{\sigma}\vec{\eta} + \pa_{\tau}\xi_1^* \pa_{\sigma}\xi_1). 
\label{vir}\end{eqnarray}

Let us devote ourselves to the type (iii) helical string solution
in $AdS_3 \times S^1$ with two spins $(S,J)$ in ref. \cite{HOS},
which is expressed in terms of a modulus parameter $q$ and Jacobi theta
and zeta functions, $\Theta_{\mu}(z,q), Z_{\mu}(z,q)$ as
\begin{eqnarray}
\eta_0 &=& \frac{C}{\sqrt{qq'}}\frac{\Theta_3(0)\Theta_0(\tilde{X}
- i\oa )}{\Theta_2(i\oa)\Theta_3(\tilde{X})}\exp \left( Z_2(i\oa)\tilde{X}
+ i\ua \tilde{T} \right), \nonumber \\
\eta_1 &=& \frac{C}{\sqrt{qq'}}\frac{\Theta_3(0)\Theta_1(\tilde{X}
- i\ob )}{\Theta_3(i\ob)\Theta_3(\tilde{X})}
\exp \left( Z_3(i\ob)\tilde{X} + i\ub \tilde{T} \right), \nonumber \\
\xi &=& \exp (i\tilde{a}\tilde{T} + i\tilde{b}\tilde{X}),
\label{etz}\end{eqnarray}
where $q' = \sqrt{1-q^2}$ and the boosted worldsheet coordinates 
$(\tilde{T},\tilde{X})$ are defined by
\begin{equation}
\tilde{T} = \frac{\mu(\tau - v \sigma)}{q'\sqrt{1-v^2}} \equiv
\frac{\ta -v \si}{\sqrt{1-v^2}}, \hspace{1cm} \tilde{X} = 
\frac{\mu(\sigma - v \tau)}{q'\sqrt{1-v^2}} \equiv
\frac{\si -v \ta}{\sqrt{1-v^2}}
\end{equation}
with $(\ta, \si)=(\mu/q')(\tau, \sigma)$. This type (iii) 
string solution with the new
modulus $q$ is produced by performing a T-transformation that is defined
by $k=iq/q'$, to the type (i) helical string solution with the
original modulus $k$. The string equations of motion for $\vec{\eta}$ in 
(\ref{st}) lead to the following relations
\begin{equation}
\ua^2 = \lu - (1-q^2)\frac{\sn^2(i\oa)}{\cn^2(i\oa)}, \hspace{1cm}
\ub^2 = \lu + \frac{1-q^2}{\dn^2(i\ob)},
\end{equation}
while the Virasoro constraints yield
\begin{eqnarray}
\tilde{a}^2 + \tilde{b}^2 &=& -q^2 - \lu - \frac{2(1-q^2)}{\cn^2(i\oa)}
+ 2\ub^2, \nonumber \\
\tilde{a}\tilde{b} &=& i C^2 \left( \frac{\ua}{q^2} \frac{\sn(i\oa)
\dn(i\oa)}{\cn^3(i\oa)} + \ub \frac{\sn(i\ob)\cn(i\ob)}{\dn^3(i\ob)}
 \right),
\label{ab}\end{eqnarray}
where the normalization constant $C$ is determined to satisfy 
$|\eta_0|^2 - |\eta_1|^2 = 1$ as
\begin{equation}
C = \left( \frac{1}{q^2\cn^2(i\oa)} + \frac{\sn^2(i\ob)}{\dn^2(i\ob)}
\right) ^{-1/2}.
\end{equation}

The period in the $\sigma$ direction of the closed string solution 
is defined from the invariance of theta functions in (\ref{etz}) as
\begin{equation}
-l \le \sigma \le l, \hspace{1cm} \Delta \sigma = \frac{2q'K(q)
\sqrt{1-v^2}}{\mu} \equiv 2l \equiv \frac{2\pi}{n}
\end{equation}
and the closedness conditions for the AdS variables are given by
\begin{eqnarray}
\Delta t &=& 2K(q)( -iZ_2(i\oa) - v\ua ) + 2n'_{\mathrm{time}}\pi
\equiv 0, \label{det} \\
\Delta \phi_1 &=& 2K(q)( -iZ_3(i\ob) - v\ub ) + (2n'_1 + 1)\pi
\equiv \frac{2\pi N_{\phi_1}}{n},
\label{dep}\end{eqnarray}
where $n = 1,2,\cdots $ counts the number of periods in 
$0 \le \sigma \le 2\pi$ and $N_{\phi_1}$ is the winding number in the
$\phi_1$ direction and $n'_{\mathrm{time}}, n'_1$ are integers.

The limit $\tilde{\omega}_{0,1} \rightarrow 0$ provides $v = 0$ from 
(\ref{det}) and generates the folded string solution from (\ref{etz}), 
which is expressed as
\begin{equation}
\eta_0 = \frac{1}{\dn(\si,q)}e^{i\ua \ta}, \hspace{1cm} 
\eta_1 = \frac{q\sn(\si,q)}{\dn(\si,q)}e^{i\ub \ta}, \hspace{1cm}
\xi_1 = \exp (i\sqrt{\lu - q^2}\ta ),
\label{fso}\end{equation}
where $\ua^2 = \lu$ and $\ub^2 = \lu + 1 - q^2$.

Now we are ready to consider the conformal SO(2,4) transformation of the
folded rotating string solution. The Polyakov action (\ref{pac}) and the
Virasoro constraints (\ref{vir}) are invariant under the isometry group
SO(2,4) of $AdS_5$. To the folded string solution (\ref{fso}) we make an
SO(2,4) transformation which is an arbitrary boost in the 
$(Y_2,Y_0)$ plane 
\begin{equation}
Y'_2 = \gamma (Y_2 + \hat{v}Y_0), \hspace{1cm} 
Y'_0 = \gamma (\hat{v}Y_2 + Y_0)
\label{abs}\end{equation}
with $\gamma = 1/\sqrt{1- \hat{v}^2}$. The transformed string 
configuration is expressed as
\begin{eqnarray}
\eta'_0 &=& \cosh \rho' e^{it'} = \frac{1}{\dn(\si)} \left[
\gamma(\hat{v}q\sn(\si) \sin\ub\ta + \cos\ua\ta ) + i \sin\ua\ta \right],
\nonumber \\ 
\eta'_1 &=& \sinh \rho' e^{i\phi'_1} = \frac{1}{\dn(\si)} \left[
q\sn(\si)\cos\ub\ta + i \gamma(q\sn(\si) \sin\ub\ta + 
\hat{v}\cos\ua\ta)\right].
\label{trs}\end{eqnarray}
When we substitute this string configuration into the string equations of
motion, $\pa_a\pa^a \vec{\eta'} - ( \pa_a\vec{\eta'}^*\cdot 
\pa^a\vec{\eta'})\vec{\eta'} = 0$, we obtain the following expression
in terms of the rescaled worldsheet variables $(\ta,\si)$
\begin{equation}
-\pa_{\ta}^2\vec{\eta'} + \pa_{\si}^2\vec{\eta'} = 
\frac{\lu \dn^2(\si) + q^2( \cn^2(\si) - \sn^2(\si) ) + q^4\sn^4(\si)}
{\dn^2(\si)} \vec{\eta'},
\label{tre}\end{equation}
which shows the invariance of $\pa_a\vec{\eta}^*\cdot \pa^a\vec{\eta}$
under the SO(2,4) transformation. The string equations of motion
are confirmed to hold for the terms including 
a $\hat{v}$ coefficient and the other terms 
separately, by taking account of $\ua^2 = \lu, \ub^2 = \lu + 1 - q^2$ 
and $\dn^2(\si)= 1- q^2\sn(\si)$. 
 
The AdS radial coordinate $\rho'$ oscillates in both $\si$ and $\ta$
such that
\begin{eqnarray}
\cosh \rho' &=& \frac{\gamma}{\dn(\si)}\left[ \left( 
\hat{v}q\sn(\si)\sin\ub\ta + \cos\ua\ta \right)^2 +
(1- \hat{v}^2)\sin^2\ua\ta \right]^{1/2}, \nonumber \\
 &=& \frac{1}{\dn(\si)}\left[ 1 + \gamma^2\left( 
\hat{v}^2q^2\sn^2(\si)\sin^2\ub\ta + 2\hat{v}q\sn(\si)\sin\ub\ta 
\cos\ua\ta + \hat{v}^2\cos^2\ua\ta \right) \right]^{1/2}, \nonumber \\
\sinh \rho' &=& \frac{1}{\dn(\si)}\left[ q^2\sn^2(\si) + \gamma^2 \left(
\hat{v}q\sn(\si)\sin\ub\ta + \cos\ua\ta \right)^2 
- \cos^2\ua\ta  \right]^{1/2},
\end{eqnarray}
which are combined to be
\begin{equation}
\cosh2\rho' = \frac{1 + q^2\sn^2(\si)}{\dn^2(\si)} 
  + \frac{2}{\dn^2(\si)} \left[\gamma^2 \left( 
\hat{v}q\sn(\si)\sin\ub\ta + \cos\ua\ta \right)^2 - \cos^2\ua\ta \right].
\label{rad}\end{equation}
The AdS time $t'$ is represented by
\begin{equation}
\tan t' = \frac{\tan\ua \ta}{\gamma \left( 1 + \hat{v}q\sn(\si)
\frac{\sin\ub\ta}{\cos\ua\ta} \right)},
\label{tim}\end{equation}
while the angular coordinate $\phi'_1$ is expressed as
\begin{equation}
\cot\phi'_1 = \frac{\cot\ub\ta}{\gamma \left( 1 + 
\frac{\hat{v}}{q\sn(\si)}\frac{\cos\ua\ta}{\sin\ub\ta} \right)}.
\label{ph}\end{equation}
From (\ref{dep}) with $n'_1= 0$, the number of periods $n$ is given by
2 for the $N_{\phi_1}=1$ case and 4 for $N_{\phi_1}=2, \cdots$, and so on
in the straight folded string. The total periodic boundary conditions 
of $t'$ and $\phi'_1$ for the closed string are satisfied owing to
$\sn(\si + 2nK(q)) = \sn(\si)$, where the total string interval is 
specified by $0 \le \si \le 2nK(q)$.

The conserved charges of the SO(2,4) transformed string configuration
are defined by
\begin{eqnarray}
E &=& \frac{\sqrt{\lambda}}{\pi}\mathcal{E} =
\frac{n\sqrt{\lambda}}{2\pi}\int_{-l}^l d\sigma \mathrm{Im}
({\eta'_0}^*\pa_{\tau}\eta'_0), \nonumber \\
S &=& \frac{\sqrt{\lambda}}{\pi}\mathcal{S} =
\frac{n\sqrt{\lambda}}{2\pi}\int_{-l}^l d\sigma \mathrm{Im}
({\eta'_1}^*\pa_{\tau}\eta'_1), \nonumber \\
J &=& \frac{\sqrt{\lambda}}{\pi}\mathcal{J} =
\frac{n\sqrt{\lambda}}{2\pi}\int_{-l}^l d\sigma \mathrm{Im}
({\xi_1}^*\pa_{\tau}\xi_1). 
\label{cch}\end{eqnarray}
By substituting the solution (\ref{trs}) into $E$ and $S$ in (\ref{cch})
the string energy is evaluated as
\begin{eqnarray}
\mathcal{E} &=& \frac{n}{2}\int_{-K(q)}^{K(q)}d\si \frac{\gamma}
{\dn^2(\si)}\left[ \ua + \hat{v}q\sn(\si)(\ua\sin\ub\tau \cos\ua\ta -
 \ub \sin\ua\ta \cos\ub\ta ) \right] \nonumber \\
 &=& \gamma \frac{n\ua}{1 - q^2}E(q)
\label{sen}\end{eqnarray}
and the AdS spin is similarly given by
\begin{eqnarray}
\mathcal{S} &=& \frac{n}{2}\int_{-K(q)}^{K(q)}d\si \frac{\gamma}
{\dn^2(\si)}\left[ q^2\ub\sn^2(\si) + \hat{v}q\sn(\si)
(\ub\sin\ub\ta \cos\ua\ta - \ua \sin\ua\ta \cos\ub\ta ) \right] 
\nonumber \\
 &=& \gamma \frac{n\ub}{1 - q^2}\left( E(q) - (1-q^2)K(q) \right).
\label{ssp}\end{eqnarray}
We see that the $\ta$-dependent terms in $\mathcal{E}$ and $\mathcal{S}$
vanish through the integration of $\sn(\si)$ over $\si$ so that
the energy and the AdS spin of the boosted string solution take
indeed constant values. The substitution of $\xi_1$ 
in (\ref{fso}) into $J$ in (\ref{cch}) yields the $S^1$ spin
\begin{equation}
\mathcal{J} = n\sqrt{\lu - q^2} K(q).
\label{js}\end{equation}
Therefore the energy-spin relation for the boosted string configuration is
given by that for the beginning folded string solution 
\cite{GKP,FT,BFS} with $E$ and $S$ replaced by $E/\gamma$ and $S/\gamma$
respectively. For instance taking a particular limit 
$\lu \approx q \ll 1$ for  (\ref{sen}), (\ref{ssp}) and (\ref{js})
we derive
\begin{equation}
E^2 = (\gamma J)^2 + \sqrt{\lambda} \gamma nS
\end{equation}
owing to $E(q) - (1-q^2)K(q) \approx \pi q^2/4$.
In the $J=0$ case which is given by $\lu = q^2, \ua = q, \ub = 1$,
the long folded string limit $q \rightarrow 1$ yields the following
energy-spin relation 
\begin{equation}
E - S = \frac{\gamma n\sqrt{\lambda}}{2\pi} \ln \left(
\frac{\pi}{\gamma n\sqrt{\lambda}} S \right),
\end{equation}
where the boost factor $\gamma$ appears together with the folding
number $n$.

Here we consider the other SO(2,4) transformation in the
$(Y_1,Y_{-1})$ plane with an arbitrary boost factor $\gamma$ 
\begin{equation}
Y'_1 = \gamma( Y_1 + \hat{v}Y_{-1} ), \hspace{1cm}
Y'_{-1} = \gamma( \hat{v}Y_1 + Y_{-1} ),
\label{bon}\end{equation}
which leads to the following string configuration 
\begin{eqnarray}
\eta'_0 &=& \frac{1}{\dn(\si)} \left[ \cos\ua\ta
 + i \gamma( \sin\ua\ta + \hat{v}q\sn(\si)\cos\ub\ta) \right],
\nonumber \\ 
\eta'_1 &=& \frac{1}{\dn(\si)} \left[\gamma(q\sn(\si) \cos\ub\ta + 
\hat{v}\sin\ua\ta) + iq\sn(\si)\sin\ub\ta \right],
\end{eqnarray}
which also obeys the string equations of motion (\ref{tre}).
The AdS radial coordinate $\rho'$ of the boosted string configuration is
characterized by
\begin{equation}
\cosh2\rho' = \frac{1 + q^2\sn^2(\si)}{\dn^2(\si)} 
  + \frac{2}{\dn^2(\si)} \left[\gamma^2 \left( \sin\ua\ta
 + \hat{v}q\sn(\si)\cos\ub\ta \right)^2 - \sin^2\ua\ta \right],
\end{equation}
which shows a similar form to (\ref{rad}).
The AdS time $t'$ and the angular coordinate $\phi'_1$ are expressed as
\begin{eqnarray}
\tan t' &=& \gamma \tan\ua\ta \left( 1 + \hat{v}q\sn(\si)
\frac{\cos\ub\ta}{\sin\ua\ta} \right), \nonumber \\
\cot \phi'_1 &=& \gamma \cot\ub\ta \left( 1 + \frac{\hat{v}}{q\sn(\si)}
\frac{\sin\ua\ta}{\cos\ub\ta} \right),
\end{eqnarray}
which are compared with(\ref{tim}) and (\ref{ph}). 
The energy and the AdS spin of this $(Y_1,Y_{-1})$ plane boosted string
solution are also described by the same expressions as 
the $\ta$-independent results in (\ref{sen})
and (\ref{ssp}) so that its energy-spin relation is the same
as that for the $(Y_2,Y_0)$ plane boosted string solution.

\section{SO(2,4) transformations of the spiky string}

Let us consider the $q \rightarrow 1$ limit configuration for 
the type (iii) helical string solution in $AdS_3 \times S^1$ (\ref{etz})
which is specified by $v = 0$ from (\ref{det}) and expressed as
\begin{equation}
\eta_0 = C\cosh(\si - i\oa)e^{i\ua \ta}, \hspace{1cm}
\eta_1 = C\sinh(\si - i\ob)e^{i\ub \ta}, \hspace{1cm}
\xi_1 = e^{i\tilde{a}\ta + i\tilde{b}\si},
\end{equation}
where 
\begin{eqnarray}
C &=& (\cos^2\ob - \sin^2\oa)^{-1/2}, \hspace{1cm}
\ua^2 = \ub^2 = \lu, \nonumber \\
\tilde{a}^2 + \tilde{b}^2 &=& -1 + \lu, \hspace{1cm}
\tilde{a}\tilde{b} = C^2(\ua\sin\oa\cos\oa + \ub\sin\ob\cos\ob).
\label{cab}\end{eqnarray}
The spikes of this string configuration reach the AdS boundary and the
energy and the AdS spin in (\ref{cch}) become divergent.

The AdS radial coordinate $\rho$ of this $q \rightarrow 1$ limit string
solution is characterized by
\begin{equation}
\cosh 2\rho = C^2\left[ (\sin^2\oa + \cos^2\ob)\sinh^2\si +
 (\cos^2\oa + \sin^2\ob)\cosh^2\si \right],
\end{equation}
from which the minimum value of $\rho$ is given by
\begin{equation}
\rho_{\mathrm{min}} = \frac{1}{2} \cosh^{-1}\left(
\frac{\cos^2\oa + \sin^2\ob }{\cos^2\ob - \sin^2\oa } \right).
\end{equation}
The AdS time $t$ and the angular coordinate $\phi_1$ are represented by
\begin{eqnarray}
\tan t &=& \frac{\tan\sqrt{\lu}\ta - \tan\oa\tanh\si}
{1 + \tan\oa \tanh\si \tan\sqrt{\lu}\ta }, \nonumber \\
\cot \phi_1 &=& \frac{\cot\sqrt{\lu}\ta + \tan\ob\coth\si}
{1 - \tan\ob \coth\si \cot\sqrt{\lu}\ta }, 
\label{ttc}\end{eqnarray}
whose expressions are compared with those in (\ref{tim}) and (\ref{ph})
for the $(Y_2,Y_0)$ plane boosted string solution.

Now we analyze the $q \rightarrow 1$ limit string configuration 
in $AdS_3$ with only AdS spin. Since it is specified 
by $\lu = 1$, that leads to 
$\tilde{a}=\tilde{b} = 0, \oa = - \ob$ through (\ref{cab}),
we express the normalization constant as $C = 1/\sqrt{\cos 2\ob}$ to have
\begin{eqnarray}
\eta_0 &=& \frac{1}{\sqrt{\cos 2\ob}}\cosh(\si + i\ob)e^{i\ta}
 \nonumber \\
 &=& \frac{1}{\sqrt{\cos 2\ob}} \Bigl[ \cos\ob\cosh\si\cos\ta -
\sin\ob\sinh\si\sin\ta \nonumber \\ 
 & & + i(\cos\ob\cosh\si\sin\ta + \sin\ob\sinh\si\cos\ta ) \Bigr], 
 \nonumber \\
\eta_1 &=& \frac{1}{\sqrt{\cos 2\ob}}\sinh(\si - i\ob)e^{i\ta}
\nonumber \\ 
 &=& \frac{1}{\sqrt{\cos 2\ob}} \Bigl[ \cos\ob\sinh\si\cos\ta +
\sin\ob\cosh\si\sin\ta \nonumber \\ 
 & & +i (\cos\ob\sinh\si\sin\ta - \sin\ob\cosh\si\cos\ta ) \Bigr].
\label{sps}\end{eqnarray}
The energy-spin relation for this
$q \rightarrow 1$ string configuration was presented \cite{HOS}
by $E - S = (n\sqrt{\lambda}/2\pi)\ln S$ which reproduces the
$n=2$ result for the straight folded rotating string solution
\cite{GKP} and the arbitrary $n$ result for the spiky string 
solution \cite{MK}.

The AdS radial coordinate of this long string solution is compactly
expressed as
\begin{equation}
\cosh 2\rho = \frac{\cosh2\si}{\cos2\ob},
\label{cor}\end{equation}
from which the minimum value of $\rho$ is given by 
\begin{equation}
\rho_{\mathrm{min}} = \frac{1}{2} \cosh^{-1}\left(
\frac{1 }{\cos 2\ob} \right).
\label{min}\end{equation}
The parameter $\ob$ is determined from (\ref{dep}) with $v=0$
and $q \rightarrow 1$. Since there are the relations among the 
Jacobi zeta functions
\begin{eqnarray}
Z_0(u+v,q) - Z_0(u,q)-Z_0(v,q) &=& -q^2 \sn(u,q)\sn(v,q)\sn(u+v,q),
\nonumber \\
Z_3(u,q) &=& Z_0(u+K(q),q),
\end{eqnarray}
the $Z_3(i\ob)$ in (\ref{dep}) is expressed as
\begin{equation}
Z_3(i\ob) = Z_0(i\ob) + Z_0(K) - \frac{\sn(i\ob)\cn(i\ob)}{\dn(i\ob)}.
\end{equation}
Using $Z_0(K) =0$ and the following asymptotic expressions in terms of
$q = e^{-r} \rightarrow 1$ 
\begin{eqnarray}
K(e^{-r}) &=& -\frac{1}{2} \ln r + \frac{3}{2}\ln 2 - \frac{1}{4}r\ln r
+ o(r\ln^mr), \nonumber \\
Z_0(i\ob,e^{-r}) &=& i\tan\ob -ir \frac{\ob + \sin\ob\cos\ob}
{2\cos^2\ob} + \frac{2i\ob}{\ln r} + O(r^2), \nonumber \\
\sn(i\ob,e^{-r}) &=& i\tan\ob + O(r), \hspace{1cm}
\cn(i\ob,e^{-r}) = \frac{1}{\cos\ob} + O(r), \nonumber \\
\dn(i\ob,e^{-r}) &=& \frac{1}{\cos\ob} + O(r),
\end{eqnarray}
which were presented in ref. \cite{HOS}, we derive
\begin{equation}
\lim_{r \rightarrow 0}K(q)Z_3(i\ob) = -i\ob.
\end{equation}
For the $N_{\phi_1}=1$ case with $n'_1 =0$ we have $2\ob = \pi - 2\pi/n$
and the real parameter $\ob$ is fixed by $\ob = \pi/6$ for $n = 3$.
Since the asymtotic expressions of $\eta_1$ at $\si = \pm K(1)$ are
given by
\begin{equation}
\eta_1(\ta,\si=K(1))= \frac{e^{K(1)}}{2\sqrt{\cos2\ob}}e^{i(\ta - \ob)},
\hspace{1cm}  \eta_1(\ta,\si=-K(1))= \frac{e^{K(1)}}{2\sqrt{\cos2\ob}}
e^{i(\ta + \ob - \pi)},
\end{equation}
the radial coordinate becomes divergent at $\si = \pm K(1)$ and the
angle difference $\Delta\phi_1 = \phi_1(\si=K(1)) -  \phi_1(\si=-K(1))$
per period is estimated as  $\Delta\phi_1 = \pi - 2\ob$.
Therefore for the $n=3$ spiky string case with one winding number 
$\Delta\phi_1$ becomes $2\pi/3$ consistently. The straight folded
string specified by $n=2, \Delta\phi_1 =\pi$ gives $\ob =0$.

Here we devote ourselves to the string configuration (\ref{sps}) for one
arc part between two spikes, which is regarded an open string that reaches
the AdS boundary. We perform the following particular two SO(2,4) boosts 
in the $(Y_2,Y_0)$ and $(Y_1,Y_{-1})$ planes to the 
$q \rightarrow 1$ limit string solution (\ref{sps})
\begin{eqnarray}
\left(\begin{array}{c}Y'_2 \\ Y'_0 \end{array} \right) &=& 
\frac{1}{\sqrt{\cos2\ob}} \left(\begin{array}{cc} \cos\ob & \sin\ob \\ 
\sin\ob & \cos\ob  \end{array} \right) \left(\begin{array}{c}Y_2 \\ Y_0 
\end{array} \right), \nonumber \\
\left(\begin{array}{c}Y'_1 \\ Y'_{-1} \end{array} \right) &=& 
\frac{1}{\sqrt{\cos2\ob}} \left(\begin{array}{cc} \cos\ob & -\sin\ob \\ 
-\sin\ob & \cos\ob  \end{array} \right) \left(\begin{array}{c}Y_1 \\ 
Y_{-1} \end{array} \right) \nonumber \\
\label{bos}\end{eqnarray}
to obtain 
\begin{eqnarray}
Y'_0 &=& \cosh\si\cos\ta, \hspace{1cm}  Y'_{-1} = \cosh\si\sin\ta, 
  \nonumber \\
Y'_1 &=& \sinh\si\cos\ta, \hspace{1cm}  Y'_2 = \sinh\si\sin\ta, 
\hspace{1cm} Y'_0Y'_2 = Y'_{-1}Y'_1,
\label{ysc}\end{eqnarray}
which give 
\begin{equation}
\eta'_0 = \cosh\si e^{i\ta}, \hspace{1cm} \eta'_1 = \sinh\si e^{i\ta}.
\end{equation}
This string solution is expressed as $\rho' = \si, 
t' = \ta, \phi'_1 = \ta$ so that it
approximates the large spin limit of the straight folded rotating 
closed string with two turning points $(n = 2)$ \cite{GKP}.
Thus the long spiky string configuration (\ref{sps}) whose shape is
characterized by the angle difference $\Delta\phi_1 = \pi - 2\ob$ is 
mapped to the long straight folded string configuration with 
$\Delta\phi_1 = \pi$ by making the double SO(2,4) boosts (\ref{bos}) 
parametrized with $\ob$.

We put $\tau \rightarrow -i\tau$ for the solution (\ref{ysc}) with 
Minkowski worldsheet in order to obtain the string solution with
Euclidean worldsheet, and interchange $Y'_2$ and $Y'_{-1}$
through $Y'_{-1} = -iY_2'', Y'_2 = -iY_{-1}'', 
Y'_1 = Y_1'', Y'_0 = Y_0''$,
which is a discrete SO(2,4) transformation, to have
\begin{eqnarray}
Y''_2 &=& \cosh\si\sinh\ta, \hspace{1cm}  Y''_{-1} = \sinh\si\sinh\ta,
  \nonumber \\
Y''_1 &=& \sinh\si\cosh\ta, \hspace{1cm}  Y''_0 = \cosh\si\cosh\ta, 
\hspace{1cm} Y_0''Y_{-1}'' = Y_2''Y_1''.
\end{eqnarray}
The resulting open string solution associated with two spikes corresponds
to the 4-cusp Wilson loop solution in the string theory computation of the
planar 4-gluon scattering amplitude \cite{AM}.
This 4-cusp Wilson loop solution is transformed by an SO(2,4) rotation
back to the elementary 1-cusp solution whose open string world surface 
ends on two semi infinite lightlike lines steming from one cusp in
the T-dual coordinates.

In ref. \cite{KAT} the SO(2,4) relation between the spiky closed string
solution with $n$ spikes and the straight folded rotating solution in
$AdS_3$ in the infinite spin limit has been argued, where the spiky
solution in the infinite spin limit is constructed as
\begin{equation}
t = \tau, \hspace{1cm} \phi_1 = \tau + \sigma, \hspace{1cm}
\cosh2\rho = \frac{\cos\sigma}{\sqrt{\sin^2\sigma_0 - \sin^2\sigma}}
\label{twr}\end{equation}
with $2\sigma_0 = 2\pi/n$, which show $\rho \rightarrow \infty$
at $\sigma \rightarrow \pm\sigma_0$ and are compared with 
 (\ref{ttc}) and (\ref{cor}).
From (\ref{twr}) the minimum value of $\rho$ is given by
\begin{equation}
\rho_{\mathrm{min}} = \frac{1}{2}\cosh^{-1}\left(
\frac{1}{\sin\sigma_0}\right),
\end{equation}
which is expressed through $\Delta\phi_1 = \pi - 2\ob = 2\sigma_0$ as
\begin{equation}
\rho_{\mathrm{min}} = \frac{1}{2}\cosh^{-1}\left(\frac{1}{\cos\ob}\right),
\end{equation}
which shows a similar value to (\ref{min}) with a slight difference.

Inversely we start with the straight folded closed string in $AdS_3$
specified by (\ref{fso}) with $\lu = q^2$ and perform a particular
SO(2,4) boost (\ref{abs}) in the $(Y_2,Y_0)$ plane with
$\hat{v} = -\tan\ob$ to have (\ref{trs}), which has the following
$q \rightarrow 1$ limit
\begin{eqnarray}
\eta'_0 &=& \frac{1}{\sqrt{\cos 2\ob}} ( - \sin\ob\sinh\si\sin\ta + 
\cos\ob\cosh\si\cos\ta ) + i\cosh\si\sin\ta = Y'_0 + iY'_{-1}, 
 \nonumber \\
\eta'_1 &=& \sinh\si\cos\ta + \frac{i}{\sqrt{\cos 2\ob}}
(\cos\ob\sinh\si\sin\ta - \sin\ob\cosh\si\cos\ta ) = Y'_1 + iY'_2.
\end{eqnarray}
To this string solution we make a special subsequent SO(2,4) 
transformation corresponding to (\ref{bon}), that is,
$Y_1'' = \gamma(Y'_1 + \hat{v}Y'_{-1}), Y_{-1}'' = \gamma(\hat{v}Y'_1
+ Y'_{-1})$ with $\hat{v} = \tan\ob$ to obtain the string
configuration specified by $\eta_0'' = Y'_0 + iY_{-1}'',
 \eta_1'' = Y_1'' + iY_2'$, which turns out to be the long spiky open
string configuration expressed by (\ref{sps}).

Here we write down the string solution which is derived by performing
these double SO(2,4) transformations to the starting folded rotating 
string solution in $AdS_3$
\begin{eqnarray}
\eta_0'' &=& \frac{\cos \ob}{\sqrt{\cos2\ob}} \frac{1}{\dn(\si)}
\left[ e^{iq\ta} + i\tan \ob q \sn(\si) e^{i\ta} \right],
 \nonumber \\
\eta_1'' &=& \frac{\cos \ob}{\sqrt{\cos2\ob}} \frac{1}{\dn(\si)}
\left[ q\sn(\si)e^{i\ta} 
- i\tan \ob e^{iq\ta}\right],
\end{eqnarray}
whose $q \rightarrow 1$ limit leads to (\ref{sps}).

\section{Conclusion}

To the straight folded rotating string solution in 
$AdS_3 \times S^1$ extracted from the type (iii) helical closed
string solution we have performed  the SO(2,4) transformation with an 
arbitrary boost parameter in the $(Y_2,Y_0)$ or $(Y_1,Y_{-1})$ plane
 to construct a new string configuration
whose AdS radial coordinate extends with the worldsheet space coordinate
and oscillates with the worldsheet time coordinate.
The SO(2,4) boosted string configuration has been confirmed to satisfy
the string equations of motion and shown to have constant values of
the energy and the AdS spin. We have demonstrated that the
energy-spin relation is derived from that of the beginning folded
rotating string solution simply by scaling the energy and the AdS spin
by the boost factor.

We have considered the shape of the long spiky open string solution
in $AdS_3$ which is extracted from the type (iii) helical closed string
solution with $n$ spikes in the string sigma model 
and analyzed how the minimum value of
AdS radial coordinate and the angle difference for one arc are
characterized by the parameter $\ob$ or $n$. By performing particular
double SO(2,4) transformations with the boosts specified
by $\ob$ in the $(Y_2,Y_0)$ and $(Y_1,Y_{-1})$ planes to the long
spiky solution  and making the analytic continuation 
together with a dicrete SO(2,4) transformation, we have
produced a long open string solution
with Euclidean worldsheet which agrees with the 4-cusp Wilson loop
solution in the computation of the 4-gluon planar amplitude.
Our expression of the long spiky string solution based on the
string sigma model is in a different form from that of ref. \cite{KAT}
where the relation between the spiky solution for the one arc part
and the Wilson loop surface solution is presented by using the
Nambu-Goto string action. But we have observed that the minimum
value of AdS radial coordinate shows a similar expression
for the two differently parametrized string solutions.

We have demonstrated that the straight folded rotating string solution
in $AdS_3$ is converted through certain two SO(2,4) rotations specified by
the boost velocities $\hat{v} = \pm\tan\ob$ with opposite signs
into a new string configuration which is expressed by the Jacobi
sn, dn functions. In the long string limit it turns out to be
the long spiky string solution whose arc shape is characterized by
$\ob$. In this inverse demonstration we have observed that the specific
boost factor $\gamma$ produces the normalization constant 
$C = 1/\sqrt{\cos2\ob}$.

\end{document}